# Mass acquisition of Dirac fermions in Cr-doped topological insulator Sb$_2$Te$_3$ films

Yeping Jiang,[1,2,3] Zhi Li,[1] Canli Song,[1,2] Mu Chen,[1,2] Richard L. Greene,[3] Ke He,[1] Lili Wang,[1] Xi Chen,[2] Xucun Ma,[1,*] and Qi-Kun Xue[2,*]

[1] *Institute of Physics, Chinese Academy of Sciences, Beijing 100190, People's Republic of China*
[2] *State Key Laboratory of Low-Dimensional Quantum Physics, Department of Physics, Tsinghua University, Beijing 100084, People's Republic of China*
[3] *Center for Nanophysics and Advanced Materials, Department of Physics, University of Maryland, College Park, Maryland 20742, USA*

We introduce time-reversal-symmetry breaking by doping Cr atoms into the topmost quintuple layer or into the bulk of Sb$_2$Te$_3$ thin films. We find that even at a high Cr-doping level the Landau level spectrum keeps a good quality, enabling the first demonstration of deviation of the zero-mode Landau level, induced by the acquisition of a mass term in the surface states in the presence of surface or bulk magnetic doping. The magnitude of the mass term in the surface states increases with increasing Cr-doping level. Our observation suggests Cr-doped Sb$_2$Te$_3$ is a promising candidate for the realization of the proposed novel magnetoelectric effects.

[*] Corresponding authors. Email: xcma@aphy.iphy.ac.cn, qkxue@mail.tsinghua.edu.cn

Three dimensional strong topological insulators (TIs), since their discovery, have attracted tremendous research interests due to the topologically nontrivial gapped bulk band and the existence of gapless Dirac fermion surface states[1, 2], which are protected by Time-reversal symmetry (TRS). The most studied three-dimensional TI systems are $Bi_2Se_3$, $Bi_2Te_3$ and $Sb_2Te_3$, which have a single Dirac cone in their surface state electronic structures within a relatively large bulk gap[3-5]. The gapless nature of topological surface states guaranteed by time-reversal symmetry (TRS) can be broken by TRS-breaking perturbations such as magnetic ordering, which changes the system into a gapped 2D Dirac fermion system where a number of novel effects based on its exotic topological magnetoelectric response (TME) have been predicted[6-8].

Ways to introduce TRS-breaking includes surface or bulk magnetic doping[9-14], and by proximity to ferromagnetic materials[7]. Although the realization of quantized TME needs the Fermi level to be tuned inside the bulk gap and the massive surface gap simultaneously, a recent theoretical work[15] suggests that one only need to tune the Fermi level inside the bulk gap. By introducing disorder to the surface, the massive Dirac fermions can be localized, while the half quantized Hall conductance $\pm \frac{e^2}{2h}$, which is essential to TME, remains due to the nontrivial bulk topology. Thus, as long as there is no change in bulk band topology[16], magnetic doping is more applicable than the proximity approach, yielding no interface but a uniform bulk gap. Experimentally, ferromagnetism does emerge in TIs with magnetic doping[17, 18]. Various results about creating a massive gap have been reported, mostly by angle-resolved photoemission spectroscopy (ARPES)[18-22]. As pointed out in a very recent work[23], several factors may contribute to the observed gap in the ARPES measurement. Scanning tunneling microscopy (STM)-based Landau level (LL) spectroscopy has been shown to be a powerful local tool to investigate the surface states dispersion[24-27]. Here we use LL spectroscopy as a mass-term sensitive probe to investigate the mass acquisition phenomena in topological surface states of Cr-doped $Sb_2Te_3$ films. We show that, both in surface- and bulk-doped samples, the zero mode in the LL spectra will deviate from the Dirac energy, indicative of a mass acquirement.

By introducing magnetic dopants, the topological surface states can be gapped with the dispersion $E_\pm = -\mu \pm \sqrt{\hbar^2 v_F^2 k^2 + m^2}$ [15]. Here $\mu$ is the chemical potential, $m = -JM_z/2\mu_B$ is the mass term coming from magnetic ordering. $M_z$ is the local magnetization in the $z$ direction. In the presence of a perpendicular magnetic field, Landau levels form in the surface state spectrum. The Landau level energies except for the zero mode are given by[28]

$$E_n = E_D \pm \sqrt{2e\hbar v_F^2 B|n| + \Delta^2}, n = \pm 1, \pm 2, \ldots, \tag{1}$$

where $\Delta = JM_z/2\mu_B + \tfrac{1}{2}g_s\mu_B B$ is the mass term (see the detailed discussion in the supplementary section).

This form of energy quantization means that, in the massive Dirac fermion system, the phase offset $\gamma$ in the Onsager quantization relation $S(E_n) = \pi k_n^2 = \pi(E_n^2 - \Delta^2)/\hbar^2 v_F^2 = 2\pi eB[n+\gamma]/\hbar$ is exactly zero, duplicating the massless case (Fig. 1a), such as graphene and TRS-preserved TIs, where the half-integer quantization is due to the Berry phase of $\pi$ [29]. In general, the quantization condition changes with Berry phase. For an ordinary electron system with parabolic dispersion and Berry phase of 0, $\gamma = \frac{1}{2}$ [30], leading to integer quantization. For gapped Dirac fermions, induced by TRS-breaking or tunneling between two copies of surface states in TIs[24, 31], or by sublattice symmetry-breaking mechanism in the graphene case, the Berry phase deviates from $\pi$ with the form $\Phi = \pi(1 - \frac{\Delta}{E})$ [7, 32]. In the TRS-breaking case, this deviation coincides with the situation that the Dirac fermions deviate from the perfect spin-momentum locking and have spin of z-component due to the out-of-plane exchange interaction. At the gap edge of surface states, the electron spin is fully polarized[28] in the perpendicular direction and has a Berry phase of 0. That the condition $\gamma = 0$ holds both in the massless and massive Dirac fermions case is due to the fact that $\gamma$ is not directly related to the exact Berry phase, but corresponds to the topological part of Berry phase[33]. This situation is unique for systems with particle-hole symmetry, which will be violated by introducing a parabolic term $\hbar^2 k^2 / 2m_e^*$ [34]. This term here is neglected due to the nearly idea linear-dispersion of surface states in $Sb_2Te_3$ films[24] compared with $Bi_2Se_3$.

The energy of zero-mode LL, which is the central issue of our work, is

$$E_0 = E_D - sgn(B) J M_z / 2\mu_B - \frac{1}{2} g_s \mu_B |B|. \tag{2}$$

Thus, the zero mode will deviate from the Dirac energy due to the mass term $\Delta$. We define the Dirac energy $E_D$ by the mid energy between ±1 LLs, as shown in Eqn. 1. The sign of Zeeman term only depends on the sign of g-factor. For $Sb_2Te_3$ films without magnetic doping, there is nearly no discernible deviation with a largest zero-mode deviation at 7 Tesla of around 1 meV, giving an upbound $g_s \approx 5$, which is relatively small and may be quite different from the bulk value. As can be seen in the experimental scanning tunneling spectrum (STS) of an undoped $Sb_2Te_3$ thin film in Fig. 1c, the zero mode position is rather symmetric relative to ±1 LL energies and nearly coincides with the energy of the state with zero density of states (DOS) at zero fields. Thus, we here only consider the mass term induced by Cr-doping. And the deviation only depends on the term $-\Delta = -sgn(B)JM_z/2\mu_B$, with the sign determined by the sign of exchange coupling $J$, the relative direction between the magnetization and the magnetic field. In Fig. 1b, we show the simulated LL spectra at 7 T for different mass term $\Delta = 0, \pm 10$ meV, respectively.

We then investigate the effect of surface or bulk Cr-doping on the topological surface states by LL spectroscopy. First we focus on the Cr-doping into the first quintuple layer (QL) of $Sb_2Te_3$ films grown by molecular beam epitaxy (MBE) (see the sample preparation section in the supplementary materials). We investigated three $Sb_2Te_3$ films with different doping levels. Two of them (sample 1 and 3) have similar Dirac energy (around 100 meV) at zero Cr-doping, while sample 2 has a lower intrinsic hole-doping ($E_D \sim$ 75 meV). During the surface Cr-doping process, the relatively low annealing temperature does not introduce extra defects other than substituting Cr into the lattice. Fig. 2a shows an STM image of Cr-doped $Sb_2Te_3$ (x=0.006). The dark triangle spot, due to the suppressed electronic states on three surface Te atoms adjacent to the substitutional Cr atom in first Sb lattice (Sb1) from the surface, is the only observed defect structure induced by Cr-substitution in the bulk doping case (see the crystal structure in Fig. 1c, the high resolution images in Fig. 2b and Fig. S4a in the supplementary section). The bright spots are intrinsic $Sb_{Te}$ (Sb on Te) defects in the topmost Te layer. We find that Cr atoms exclusively substitute for Sb atoms probably due to the large difference in ion sizes between Cr and Te atoms. The surface deposition and annealing procedure only introduces $Cr_{Sb1}$ demonstrated by the similarity between the number of surface Cr atoms before annealing and that of $Cr_{Sb1}$ defects after annealing. Invisible $Cr_{Sb2}$ in the bulk-doping case is due to the fact that Cr-doping in $Sb_2Te_3$ ($Cr_{Sb}$) is isovalent with no charge-doping [35], in contrast to the case of intrinsic defects where the defect structures can be imaged by STM even for defects in the fifth atomic layer from the surface[36]. The clover-shaped defect structures of intrinsic defects are essentially the effect of defect-induced local band-bending, especially in the case of Sb vacancy.

For the lowest doping case (x=0.006), no apparent zero-mode deviation can be observed. We increased the Cr-doping stepwise and observed the zero-mode deviation at each doping level. As can be seen from the atomic resolution STM image (Fig. 2d), even at a high doping level (x=0.18), the Cr-doped first QL is well ordered and exhibits a clean surface. The Cr-doping is quite uniform in the length scale of 10 nm. The nearly undisturbed crystal lattice, the clean surface, the isovalent and uniform nature of doping, all together make it possible for the LL spectra to keep a good quality and enable the possibility of observing the zero-mode behavior by, other than introducing a mass term to the surface states, not inducing considerable scattering events. Fig. 2e shows a corresponding LL spectrum taken on the $Sb_2Te_3$ film with

x=0.18, where a series of well-defined LL peaks can be seen. Compared with the undoped case (Fig. 1c) where the TI surface states are expected massless[24], the well separated zero mode is no longer symmetric relative to the positions of ±1 LLs, shifting positively with a magnitude of about 6 meV from the Dirac energy, implying the existence of a mass term. We assume that at a high magnetic field of 7 T, the z-components of magnetic moments align with the magnetic field in the same direction. The positive deviation (negative mass term $\Delta$) implies an antiferromagnetic exchange interaction $J$ between the spin of surface state electrons and that of the Cr dopants.

Besides the zero-mode deviation, the Cr-doping lifts up the Dirac energy. Fig. 2f shows LL spectra at 7 T with the increasing Cr-doping levels. The Dirac energy moves steadily upwards, from around 100 meV without doping to around 170 meV in the heavily doped case (x=0.21). This is not caused by a carrier-doping effect. The Fermi level is nearly unchanged relative to the bulk-like valence band edge (dashed line). We observed a similar behavior in the bulk Cr-doping case (see Fig. S4c in the supplementary section), ruling out the possibility of surface band-bending due to the surface doping. This is a rigid band shift between the surface Dirac cone and the bulk-like band, supported by the increasing or decreasing number of LLs below or above the Dirac energy. The shaded region, which does not change with Cr-doping, is roughly the bulk gap where LLs emerge. While the existence of Dirac point (Kramer degenerate point) is guaranteed by TRS, its energy varies with boundary conditions. Theoretically, the Dirac cone can be engineered by changing the elements in the topmost atomic layer[37]. Experimentally, the Dirac cone engineering has been achieved by bulk-doping in the $(Bi_{1-x}Sb_x)_2Te_3$ case[38]. We see similar behavior in the bulk Cr-doping case, as will be shown later.

From the aforementioned Onsager quantization condition $k_n^2 = 2\frac{e}{\hbar}|n|B$ and Eqn. 1, the dispersion of gapped surface states can be simulated by taking field dependent LL spectra. In Fig. 3 we show the data in three samples with different Cr-doping levels: sample 3 (x=0.105), sample 2 (x=0.18), sample 1 (x=0.21). On the film of x=0.105 (Figs. 3a and 3c), the LLs energies except the zero mode keep a good linearity, just like the case without Cr-doping. Note that the deviation for LLs with higher $|n|B$ from the gapless case is nearly indiscernible (less than 0.5 meV for $\Delta$=5 meV and lowest available $|n|B$: n=1, B=3 T). For the x=0.105 case, the zero mode energy gradually deviates from the Dirac energy and reaches the high-field value above 2 T, indicating a field-dependent behavior of mass term $\Delta(B)$ or magnetization $M_z(B)$. For films at even higher Cr-doping levels, the zero mode deviation is nearly constant above the magnetic field at which it becomes discernible, implying a low saturation field of ≤ 1 T for magnetization of dopants in the first QL. The fitting of the data to Eqn. 1 is quite good (Fig. 3d), showing a massive Dirac dispersion. The field dependent data is similar for the film with even higher Cr-doping (x=0.21), which exhibits a relatively larger massive gap as shown in Fig. 3e.

Remembering that the sign of mass term depends on the relative direction between the magnetization and magnetic field[28], the zero-mode may flip from above $E_D$ to below $E_D$ by flipping the magnetic field into opposite direction but with a magnitude smaller than the coercive field of possible ferromagnetism in the surface-doping case (see Fig. S3a in the supplementary section). We measured the surface state spectra by changing the magnetic field in small steps around zero (see Fig. S3b in the supplementary section). The zero-mode flipping behavior was not observed. Actually, in order to see this behavior, a coercive field of around 1 T is needed (zero mode does not appear below 1 T), while for this material it's only about 0.1 T[35]. Thus, the zero-field mass term or ferromagnetism in the surface-doping case remains open for further study.

In Fig. 4, we summarize the results of Cr-doping effect on the topological surface states in $Sb_2Te_3$ films, as illustrated by the schematic in Fig. 4a. First, the Dirac cone shifts rigidly upwards relative to the bulk-like band. Fig. 4b is the doping dependent Dirac energy for two different intrinsic doped samples, showing the upward shift in $E_D$ of about 70 meV at x=0.21. The offset in energy between the two data sets originates from different Fermi energies, which seldom change with the increasing Cr-doping. Second, the mass term $\Delta$ or magnetization of dopants $M_z$ increases with Cr-doping level in a monotonous manner, reaching a value of about 10 meV at x=0.21 (Fig. 4c), comparable to the theoretical value[9]. In addition, there is nearly no difference between samples of the same Cr-doping level but with different intrinsic doping or chemical potential. The error bar results from the local inhomogeneity in Cr-doping. In Fig. S3c of supplementary section, we show the spatial variation in zero-mode deviation (2.4~7.2 meV) for doping level x=0.12. From the varying zero-mode positions and the unvarying ±1 LLs, we can see clearly the irrelevance between variation in the zero-mode deviation and charge doping.

Finally, for comparison, we discuss the bulk Cr-doping case. The bulk-doping also lifts the Dirac energy and introduces a mass term, but with much larger magnitudes. The Dirac energies shift to approximately 160 meV and 180 meV for x=0.03 and 0.08, with the corresponding mass terms $\Delta$ at 7 T of around 11 meV and 16 meV, respectively. For surface doping of x=0.08, $\Delta$ is about 4 meV, about 4 times smaller in magnitude than that of bulk-doping (see Figs. 4b and 4c, or Figs. S4c and S4e in the Supplementary Information for more details). It is understandable considering that in our surface-doping scheme Cr atoms only substitute the first Sb layer. Based on the observation that the critical thickness for $Sb_2Te_3$ being a three dimensional TI is around 4 QL[24], the surface state electrons are supposed to have a spatial extension of about 2 QLs, implying that the Dirac fermions may have interaction with the dopants' spins in the first four Sb atomic layers. In addition, it shows a high homogeneity in the spatial distribution of mass term, probably due to a more uniform overall Cr-doping level in 2 QLs than that just in the first Sb layer. Besides, the zero-field spectra at higher bulk Cr-doping show more pronounced gapped features after turning on and turning off a perpendicular magnetic field (Fig. S4f in the supplementary section). Thus, although the zero-mode deviation demonstrates the existence of a mass term for the Dirac fermions in a magnetic field both in the surface- and bulk-doping cases, compared with the surface-doping case where there seems no clear signature of mass acquisition at zero field, there is evidence of easier spontaneous perpendicular alignment of magnetic moments in

Sb$_2$Te$_3$ films with bulk Cr-doping. The introducing of a mass term, the isovalent nature of doping, and the effect of Dirac energy lifting imply that the Cr doped Sb$_2$Te$_3$ film is a potential candidate of realizing many exotic topological magnetoelectric phenomena.

**Acknowledgements:** This work was supported by National Science Foundation and Ministry of Science and Technology of China. Y. P. Jiang was partly supported by the NSF-USA (grant DMR-1104256).

**Figure captions:**

FIG. 1 (color online). (a) Schematic of surface states of a 3D TI and the Landau quantization scheme with and without massive gap neglecting the field induced Zeeman term. The sign of zero-mode deviation depends on the sign of TRS-breaking mass term. (b) Simulation of Landau level formation (we take similar experimental values of peak widths and Fermi velocity) in a perpendicular magnetic field of 7 T with mass terms of $\Delta = 0, \pm 10$ meV. (c) The STS on an $Sb_2Te_3$ film (sample 3 before Cr-doping) at 0 T and 7 T. Tunneling gap conditions: $V_{bias}$ = 0.2 V, I = 200 pA. All STM data was taken at 4.8 K. In the STS measurement, the bias modulation was 1 mV at 987.5 Hz. The bias voltage was applied to the sample.

FIG. 2 (color online). (a) STM image (-1.0 V, I = 50 pA) of a Cr-doped $Sb_2Te_3$ film (x=0.006). Here 'x' denotes the ratio between Cr-doping in the first QL and surface atom density. (b) Zoom-in high-resolution image (0.1 V, 40 pA) of $Cr_{Sb1}$ defects. (c) Schematic of surface Cr-doping and the atomic structure of 1 QL $Sb_2Te_3$, showing Cr on Sb1 (first Sb layer from the surface) and Sb2. (d) (e) High resolution STM image (0.2 V, 400 pA) and STS (0.2 V, 200 pA) at 7 T of sample (x=0.18). (f) STS at 7 T (0.2 V, 200 pA) on sample 1 with increasing Cr-doping x: 0, 0.03, 0.06, 0.09, 0.12, 0.18, 0.21. The arrows indicate the zero modes. The dashed short lines indicate the ±1 LLs, while the solid short ones indicate the Dirac energies. Spectra of different doping are shifted vertically for clarity.

FIG. 3 (color online). (a) Field dependent LL spectra (0.2 V, 200 pA) of different samples with

x=0.105 (sample 3) and 0.18 (sample 2). The solid line indicates the zero modes at 7 T. (c)-(e) LL energies under various magnetic fields plotted versus $sgn(n)\sqrt{|n|B}$ on samples of x=0.105 (sample 3), 0.18 (sample 2), 0.21 (sample 1), respectively. The red line in (c) is a linear fitting to the date except the zero modes. The dashed curves in (d)(e) are the fitting according to the quantization condition $E_n = sgn(n)\sqrt{2e\hbar v_F^2 B|n| + \Delta^2}$ of the gapped surface states.

FIG. 4 (color online). (a) Schematic of the effect of Cr-doping on the surface state structure. Here $E_D$ indicates Dirac energy of surface states, which can be obtained from the ±1 LL energies. CBM and VBM indicate the band edge of bulk-like conduction band (BCB) and valence band (BVB). (b)(c) Doping dependence of Dirac energy $E_D$ and the mass term $\Delta$. (d) Field-dependent mass term $\Delta(B)$ of samples with different Cr-doping levels.

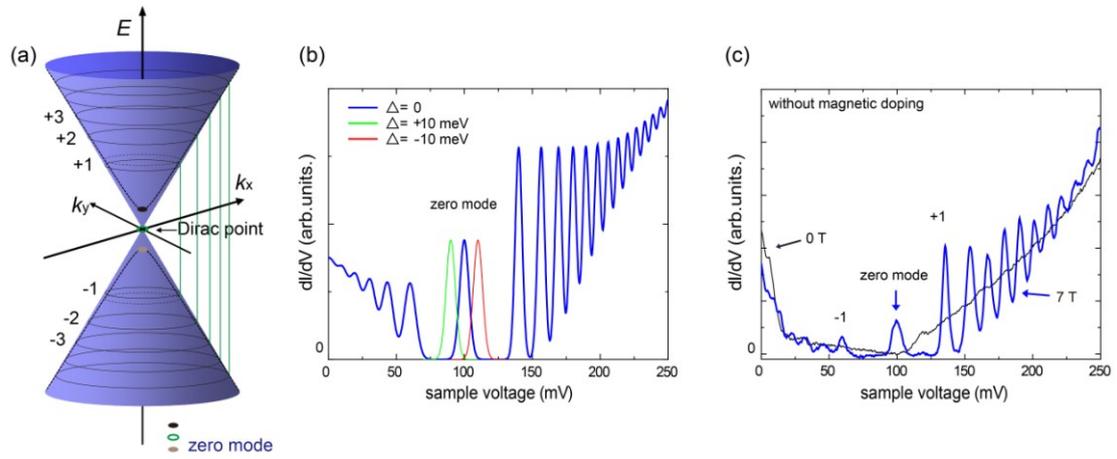

**Figure 1**

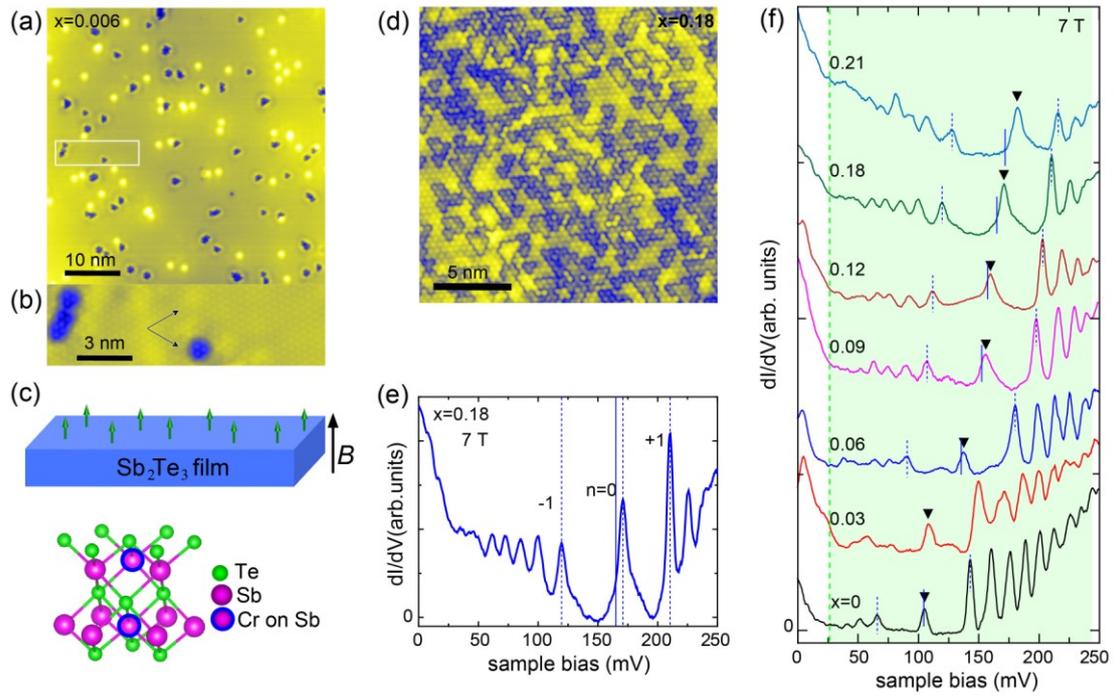

**Figure 2**

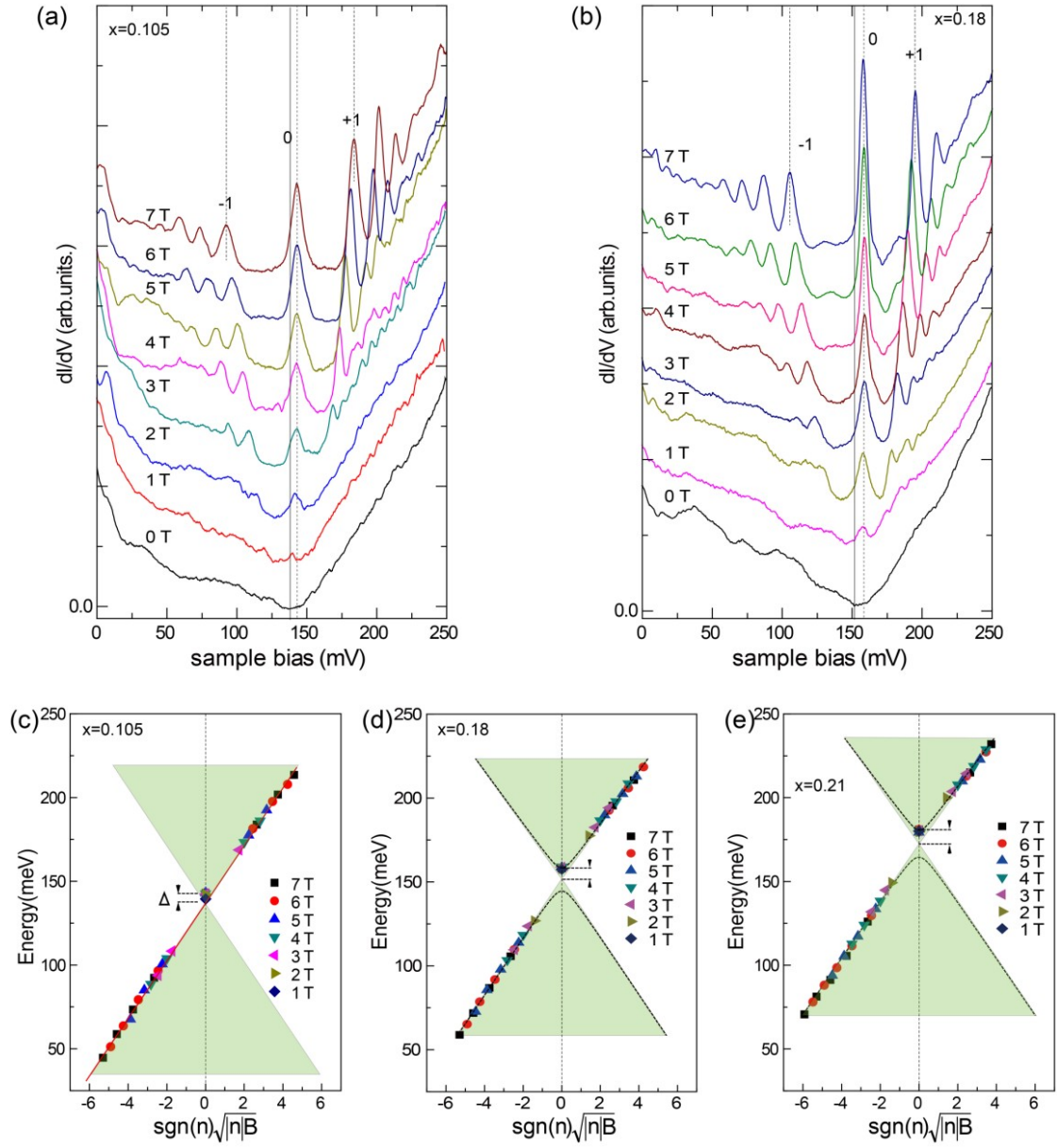

**Figure 3**

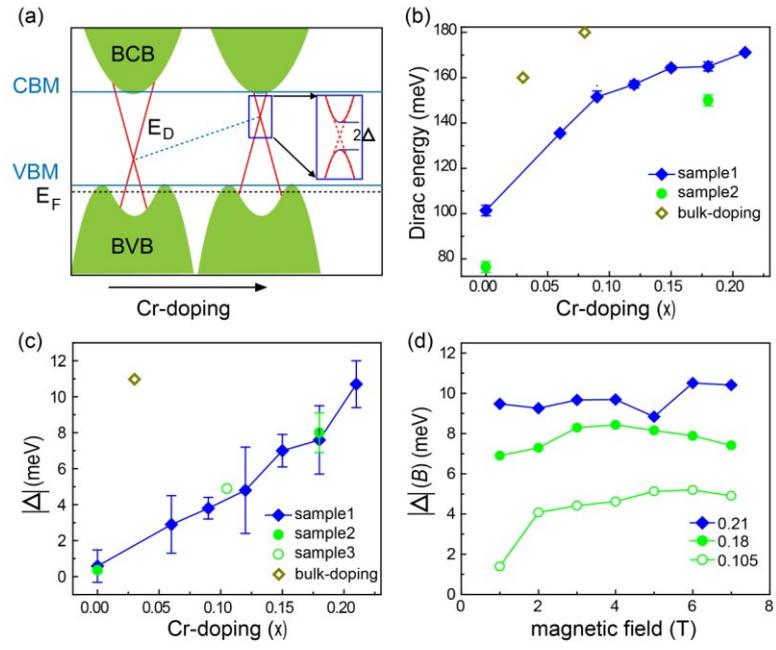

**Figure 4**

# Supplementary information for: Mass acquisition of Dirac fermions in Cr doped topological insulator $Sb_2Te_3$ films

Yeping Jiang,[1,2,3] Zhi Li,[1] Canli Song,[1,2] Mu Chen,[1,2] Richard L. Greene,[3] Ke He,[1] Lili Wang,[1] Xi Chen,[2] Xucun Ma,[1,]* and Qi-Kun Xue[2,]*

[1] *Institute of Physics, Chinese Academy of Sciences, Beijing 100190, People's Republic of China*
[2] *State Key Laboratory of Low-Dimensional Quantum Physics, Department of Physics, Tsinghua University, Beijing 100084, People's Republic of China*
[3] *Center for Nanophysics and Advanced Materials, Department of Physics, University of Maryland, College Park, Maryland 20742, USA*

*email: xcma@aphy.iphy.ac.cn; qkxue@mail.tsinghua.edu.cn

**I. Mass term accquisition and Landau level formation in the presence of magnetic doping and the perpendicular magnetic field.**

In the presence of magnetic dopants, the Dirac Hamiltonian of topological surface states is $H = -\mu I + \hbar v_F(k_x\sigma_2 - k_y\sigma_1) + m\sigma_3$, where a mass $m$ comes into play because of the exchange interaction between magnetic dopants and surface state electrons: $H_{ex} = -J\sum S_i \cdot s\delta(r - R_i)$. By only taking the effect of the z-component spin (perpendicular direction) of dopants, $H_{ex}|_z = -Jn_{Cr}\bar{S}_z\sigma_3/2$. Here $\mu$ is the chemical potential relative to the charge neutrality point (Dirac point $E_D = -\mu$), $\sigma_i$ are spin Pauli matrices, $\bar{S}_z$ is the z-component of averaged dopant spin, $n_{Cr}$ is two dimensional doping density. Thus $m = -\frac{1}{2}Jn_{Cr}\bar{S}_z = -JM_z/2\mu_B$, where $M_z$ is the local magnetization in the z direction. The in-plane spin components, though also being TRS-breaking terms, only shift the Dirac point in the equi-energy plane in momentum space, and are not included here. The resulting surface state dispersion is $E_\pm = -\mu \pm \sqrt{\hbar^2 v_F^2 k^2 + m^2}$, with a gap of $2m$ in the surface states.

In the presence of a perpendicular magnetic field, Landau levels form in the surface state spectrum. The Hamiltonian in the field simply becomes $H = -\mu I + \hbar v_F(k'_x\sigma_2 - k'_y\sigma_1) + (m + \frac{1}{2}g_s\mu_B B)\sigma_3$, taking into account the effect of magnetic field by adding the magnetic vector potential to the momentum operator $\hbar k' = \hbar k + eA$ and adding the Zeeman term $\frac{1}{2}g_s\mu_B B\sigma_3$, where $g_s$ is the g-factor of surface state electrons. Similar to exchange term, in the Zeeman term $H_z = g_s\mu_B\hbar^{-1}\mathbf{B}\cdot\mathbf{s}$, only a perpendicular magnetic field will add a mass term to the surface states. The Landau level energies except for the zero mode are given by

$$E_n = E_D \pm \sqrt{2e\hbar v_F^2 B|n| + \Delta^2}, n = \pm 1, \pm 2, \ldots,$$

where $\Delta = J M_z / 2\mu_B + \frac{1}{2} g_s \mu_B B$ is the mass term. The energy of zero-mode LL is

$$E_0 = E_D - sgn(B) J M_z / 2\mu_B - \frac{1}{2} g_s \mu_B |B|.$$

**II. Sample preparation.**

All experiments were carried out in an ultrahigh-vacuum system, equipped with an MBE and a low temperature STM having a 7 Tesla magnet. The base pressure of the system is better than $1 \times 10^{-10}$ torr. In the MBE growth of $Sb_2Te_3$ films, highly n-doped and graphitized 6H-SiC(0001) covered mainly by bilayer graphene was used as the substrate. $Sb_2Te_3$ films were prepared by thermal evaporation of high-purity Sb (99.9999%) and Te (99.9999%) from two standard Knudsen cells. During growth, the temperatures of the Sb source, the Te source, and the substrate were set at around 330 °C, 225 °C, and 230 °C, respectively, resulting in a growth rate of about 0.2 QL/min and a Te/Sb flux ratio of about 10. By varying the growth conditions films with different intrinsic doping, sample 1 (~50 QL), sample 2 (~40 QL), and sample 3 (~50 QL), were obtained. The Cr-doping into the first Sb atomic layer was achieved by low-temperature (100 K) deposition of Cr and annealing at 200 °C for hours, while the bulk Cr-doping was carried out by co-deposition of Sb, Te, and Cr. High-purity Cr atoms (99.999%) were evaporated from a tantalum boat by electric current heating (~1000 °C). In the process of Cr-doping into the first Sb atomic layer, a stepwise strategy, with a maximum doping of x=0.03 in each step, was used to avoid Cr-cluster formation.

**III. Supplementary figures.**

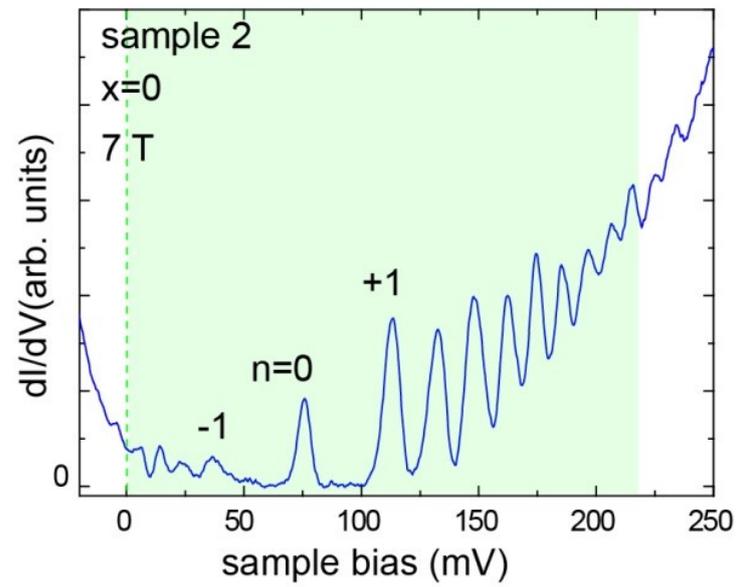

**Figure S1 | LL spectrum of sample 2 without Cr-doping.** Consistent with Fig. 3b, the Fermi level in sample 2 lies nearly at the bulk-like valence band edge where DOS begins to increase sharply.

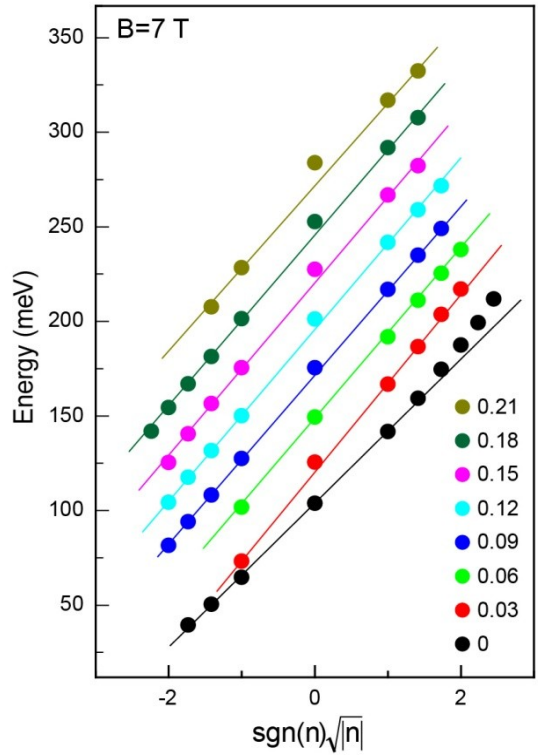

**Figure S2 | Doping-dependence of zero-mode deviation.** Energies of Landau levels were plotted against $sgn(n)\sqrt{|n|}$ at various Cr-doping levels in sample 1. The linear fitting to the data except for zero mode gives a Fermi velocity of $4.8\times10^{-5}$ m/s. In Fig. 4c, the data for x=0.03 is not shown due to the low quality of LL spectrum at low Cr-doping as shown in Fig. 2f. The large zero-mode deviation at this doping level in the figure here is due to the uncertainty in determining the peak position.

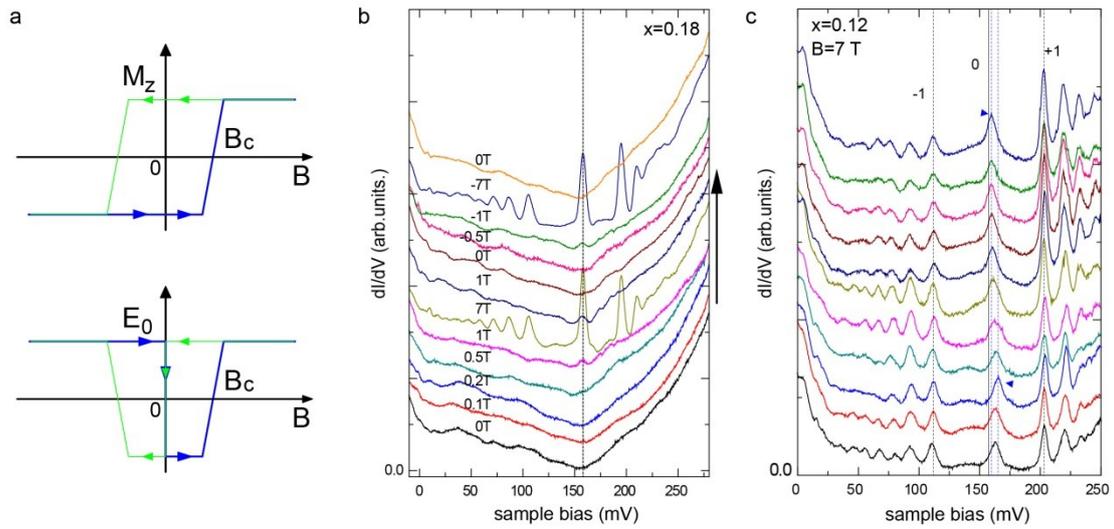

**Figure S3 | Absence of zero-mode flipping and site-dependence of zero-mode deviation. a**, The two diagrams illustrate the hysteresis loop of possible ferromagnetism and the corresponding zero-mode flipping in a varying magnetic field, respectively. **b**, Surface state spectra at various magnetic fields swept in small steps and in opposite directions. The Cr-doping level is x=0.18 (sample 2). The arrow indicates the sweeping direction of the magnetic fields. **c**, Landau level spectra taken at random positions on an $Sb_2Te_3$ film (sample 1) of Cr-doping x=0.12, showing the spatial variation of zero-mode deviation (2.4 ~ 7.2 meV). The arrows indicate the zero modes coinciding with the dashed lines. The solid line and dashed lines around the zero mode indicate the Dirac energy and the range of zero-mode deviations, respectively. We see that the zero mode always deviates in the positive directions.

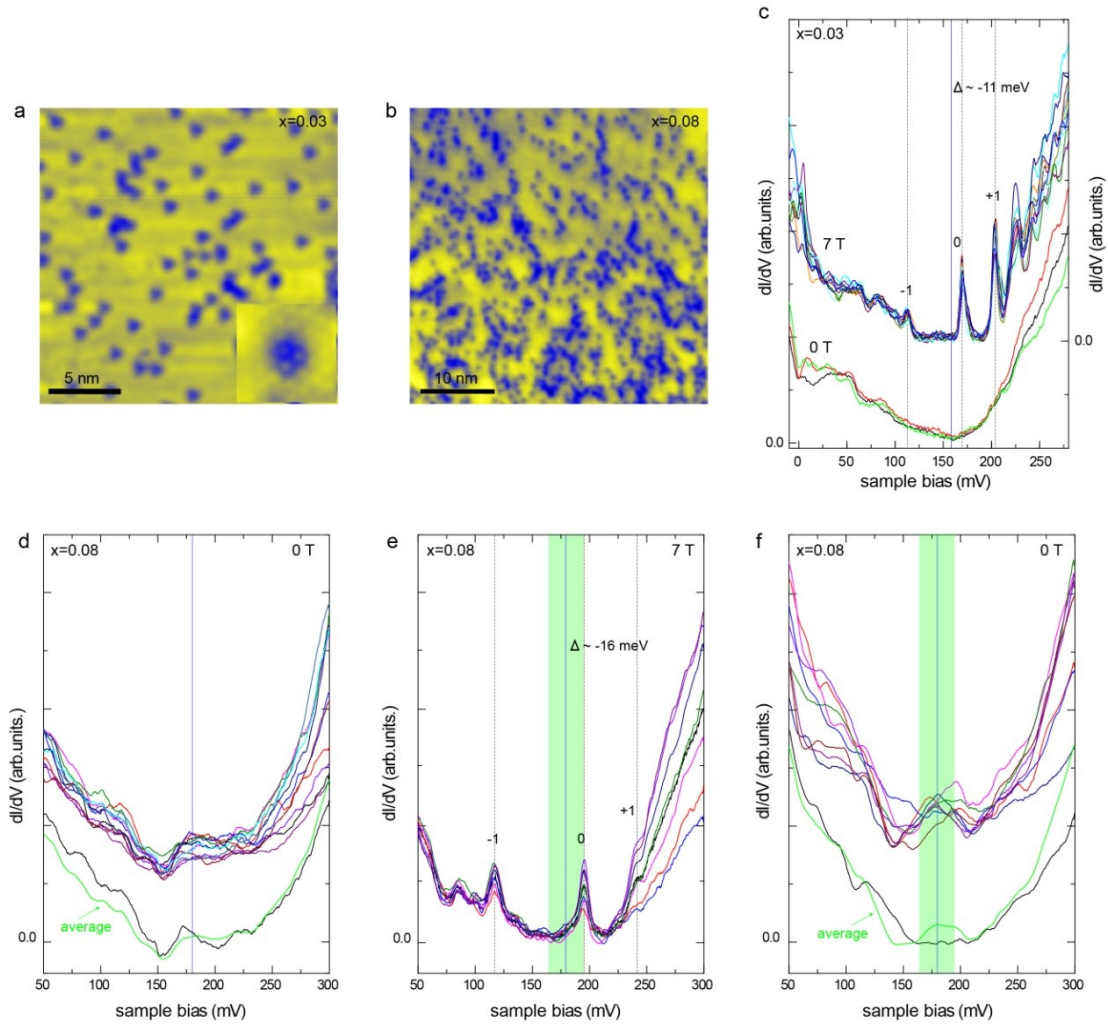

**Figure S4 | Zero-mode deviation in samples with bulk Cr-doping and the possible gapped surface states at zero field. a**, **b**, STM images (-1.0 V, 200 pA for **a**; 1.0 V, 50 pA for **b**) of $Sb_{2(1-x)}Cr_{2x}Te_3$ (bulk doping) surfaces. The inset in **a** shows the atomic resolution image of dark dots induced by Cr-doping. **c**, Zero-field STS and Landau level spectra (0.2 V, 200 pA) taken at random positions on an $Sb_{2(1-x)}Cr_{2x}Te_3$ (bulk doping, x=0.03) film. The LL spectra are shifted vertically for clarity. **d**, **e**, **f**, Surface state spectra (0.25 V, 200 pA) taken at random positions on $Sb_{2(1-x)}Cr_{2x}Te_3$ (bulk doping, x=0.08) before applying the magnetic field, after turning on the magnetic field and after turning off the magnetic field. The +1 LL in **e** is barely resolved because

it becomes quite close to the bulk-like conduction band edge due to the lifted Dirac energy. The green curves in **d**, **f** are spatially averaged curves. The black curves are curves at particular positions. The blue solid lines indicate the Dirac energy, while the green shadowed region demonstrates the possible surface gap implied by the zero-mode deviation as shown in **e**. The peaked feature in the surface state gap may be due to the impurity states[1]. Applying and removing a magnetic field may enhance the ferromagnetism and make the surface state gap more uniform. Thus the peak becomes more pronounced because of suppressed surface state DOS around the Dirac point.